\documentclass[12pt]{iopart}
\usepackage{iopams}
\usepackage{latexsym}
\usepackage{amssymb}
\usepackage{exscale}
\usepackage{bm}
\usepackage{graphicx}
\usepackage{setstack}
\newcommand{\tensor}[1]{{\mathsf {#1}}}
\newcommand{\Nabla}{\boldsymbol{\nabla}}
\newcommand{\Lnabla}{\overset{\leftarrow}{\Nabla}}

\newcommand{\D}{\mathrm{d}}

\newcommand{\fo}[1]{\underline{\hat{\mathbf{#1}}}}

\begin{document}
\title[Matter-screened
Casimir force and Casimir-Polder force]{Matter-screened
Casimir force and Casimir-Polder force in planar structures}
\author{Christian Raabe and Dirk-Gunnar Welsch}
\address{Theoretisch-Physikalisches Institut,
Friedrich-Schiller-Universit\"at Jena, Max-Wien-Platz 1, D-07743 Jena,
Germany}
\ead{C.Raabe@tpi.uni-jena.de, D.-G.Welsch@tpi.uni-jena.de}
\date{\today}

%%%%%%%%%%%%%%%%%%%%%%%%%%%%%%%%%%%%%%%%%%%%%%%%%%%%%%%%%%%%%%%%%%%%%%%%%%%
%\submitto{\JOB}
%%%%%%%%%%%%%%%%%%%%%%%%%%%%%%%%%%%%%%%%%%%%%%%%%%%%%%%%%%%%%%%%%%%%%%%%%%%
\begin{abstract}
Using a recently developed theory of the Casimir
force (Raabe C and Welsch D-G 2005 \emph{Phys. Rev.} A
{\bfseries 71} 013814),
we calculate the force that acts on a plate in front
of a planar wall and the force that acts on the plate
in the case where the plate is part of matter that fills
the space in front of the wall. We show that in the
limit of a dielectric plate whose permittivity is close to unity,
the force obtained in the former case reduces to the
ordinary, i.e., unscreened Casimir-Polder force acting
on isolated atoms. In the latter case, the theory
yields the Casimir-Polder force that is screened
by the surrounding matter.
%proof:
\\[1ex]
\end{abstract}
%
%\submitto{\JOB}
%%%%%%%%%%%%%%%%%%%%%%%%%%%%%%%%%%%%%%%%%%%%%%%%%%%%%%%%%%%%%%%%%%%%%%%%%%%
%proof: no pacs
%\pacs{
%%%03.70.+k, %Theory of quantized fields 
%12.20.-m, %Quantum electrodynamics
%%%12.20.Ds, %Specific calculations
%32.80 .Lg, %Mechanical effects of light on atoms, molecules, and ions
%42.50.-p, %Quantum optics
%42.50.Vk, %Mechanical effects of light on atoms, molecules, electrons,
%%and ions
%42.50.Nn %Quantum optical phenomena in absorbing, dispersive
%%and conducting media
%%
%%%42.60.Da, %resonators, cavities, amplifiers, arrays and rings
%}
%
%%%%%%%%%%%%%%%%%%%%%%%%%%%%%%%%%%%%%%%%%%%%%%%%%%%%%%%%%%%%%%%%%%%%%%%%%%%%
\maketitle
%%%%%%%%%%%%%%%%%%%%%%%%%%%%%%%%%%%%%%%%%%%%%%%%%%%%%%%%%%%%%%%%%%%%%%%%%%%%
\section{Introduction}
\label{sec1}

%%%%%%%%%%%%%%%%%%%%%%%%%%%%%%%%%%%%%%%%%%%%%%%%%%%%%%%%%%%%%%%%%%%%%%%%%%%%
%
From the point of view of statistical physics, the
classical electromagnetic vacuum
can be characterized by the condition that
all moments of the electric and induction
fields vanish identically, which implies the
absence of any interaction with matter.
In quantum electrodynamics, this definition
is precluded by the non-commutativity
of canonically conjugate field quantities;
thus non-vanishing moments must inevitably occur.
The quantum electromagnetic vacuum can be
merely characterized as the state in which
all normally ordered field moments
vanish identically. Clearly, the anti-normally
(or otherwise non-normally) ordered
field moments then cannot do so, due to
virtual photon creation and destruction
-- a signature of the noise of the quantum vacuum.
Since the electromagnetic vacuum cannot be switched
off, its interaction with atomic systems cannot
be switched off either, thereby giving rise to
a number of observable effects.
Both virtual and real photons can be involved in the
interaction. Whereas the interaction of ground-state
atoms with the electromagnetic vacuum takes place
via virtual photon creation and destruction,
the creation of real photons always requires
excited atoms.

A typical example of the
interaction via virtual photons
is the attractive van der Waals force
between two unpolarized ground-state atoms,
which can be regarded as the force between
electric dipoles that are induced by the
fluctuating vacuum field. In the non-retarded
(i.e., short-distance) limit,
the potential associated with the force
%proof:
%has been
was
first calculated by London
\cite{LondonF001930,LondonF001937}. The theory
was later extended by Casimir and Polder
\cite{CasimirHBG021948} to allow for larger separations,
where retardation effects cannot be disregarded.
Forces which are mediated by the electromagnetic
vacuum are not only observed on a microscopic level
but also on macroscopic levels. Typical examples
are the force that an (unpolarized) atom experiences
in the presence of macroscopic (unpolarized) bodies
-- referred to as Casimir-Polder (CP) force
in the following -- or the Casimir force
between macroscopic (unpolarized) bodies
(for a review, see,
%proof:
%e.g., reference
for example, %
\cite{MilonniQuantumVacuum}).

Since macroscopic bodies consist of a huge number of
atoms, both the CP force and the Casimir force
can be regarded as macroscopic manifestations of
microscopic van der Waals forces, and both types of forces
are intimately related to each other.
Clearly, they cannot be obtained, in general, from
a simple superposition of two-atom van der
Waals forces because such a procedure would
completely ignore
the interaction between the constituent atoms of the
bodies, and thus also their collective influence on the
structure of the body-assisted electromagnetic field
\cite{LifshitzEM001955}.
The aim of the present paper is to study
this problem in more detail, with
special emphasis on the CP force in
planar structures.

In principle, it is certainly possible to calculate
CP and Casimir forces within the framework of
microscopic quantum electrodynamics, by solving
the respective many-particle problem in some approximation.
Alternatively, one can start from a macroscopic description
of the bodies in terms of boundary conditions
or, more generally, in terms of polarization and magnetization
fields together with (phenomenologically introduced)
constitutive relations.
The latter, very powerful
approach will be used throughout this paper.
To be more specific, we will apply the
theory recently developed in
%proof:
%reference
%
\cite{RaabeC012005},
which renders it possible not only to calculate the
Casimir force that acts on bodies separated by
empty space, but also the one which acts on bodies
the interspace between which is filled with matter.
As we will see, the formula for the Casimir force
obtained in this way contains,
as a special case, the well-known formula for the CP force acting on
isolated atoms. Moreover, it can also be used to calculate the
CP force acting on atoms that are constituents of matter,
where the neighbouring atoms give rise to a screening effect
that diminishes the force.
Throughout the paper systems that are at rest are
considered, which implies that the electromagnetic
vacuum forces must be thought of as being
balanced by some other forces.

The paper is organized as follows.
After a
%proof:
%recall
review
in section \ref{sec2} of
the CP force, the theory of the Casimir force as
developed in
%proof:
%reference
\cite{RaabeC012005} is outlined
in section \ref{sec3}, and the Casimir stress in
planar structures is given.
Relations between the Casimir force and the unscreened as well as
the screened CP force are studied in section \ref{sec4}, and the
results are discussed and summarized
in section \ref{sec5}.

\section{Casimir-Polder force}
\label{sec2}
Provided that the broadening of the atomic levels can be
neglected, the CP force is conservative, i.e., expressible as the
(negative) gradient of a potential -- the CP potential.
While this approximation
may be invalid for atoms prepared in an excited state, it is
well justified for ground-state atoms,
in which case the CP potential
can be written in the form of
\begin{equation}
\label{L6}
V^\mathrm{(at)}(\mathbf{r})=\frac{\hbar\mu_{0}}{2\pi}
\int_{0}^{\infty}\D\xi\,
\xi^2 \alpha(i\xi) \mathrm{Tr\,}\tensor{G}^\mathrm{(S)}
(\mathbf{r},\mathbf{r},i\xi)
%.
\end{equation}
(for derivations, see,
%proof:
%e.g., references
for example,
\cite{McLachlanAD001963,AgarwalGS001975,
WylieJM001984,HenkelC002002,BuhmannSY032004}).
Here, $\mathbf{r}$ is the position of the atom,
$\alpha(i\xi)$ is its (ground-state)
polarizability and $\tensor{G}^\mathrm{(S)}(\mathbf{r,r'},i\xi)$
is the scattering part of the classical retarded Green tensor
$\tensor{G}(\mathbf{r,r'},i\xi)$
on the imaginary frequency axis.
Note that the Green tensor takes the presence of arbitrary (locally
responding) magnetodielectric bodies
into account within the framework of macroscopic
linear electrodynamics in causal media.
Only the scattering part
of the Green tensor figures in equation (\ref{L6}); its bulk part,
despite being divergent in the
coincidence limit $\mathbf{r'}$ $\!\to$ $\!\mathbf{r}$,
does not contribute to the force on the atom
and can be thus discarded from equation (\ref{L6}).
The potential (\ref{L6}), which is valid to first order
in $\alpha(i\xi)$, can be derived
using the Green tensor scheme of electromagnetic field
quantization \cite{KnoellL002001}
and treating the atom--field interaction in
the electric dipole approximation and the lowest
(non-vanishing) order of perturbation theory.

In particular, for a ground-state atom
in front of a planar magnetodielectric wall, equation
(\ref{L6}) leads to (see,
%proof:
%e.g., references
for example, %
\cite{McLachlanAD001963,AgarwalGS001975,WylieJM001984,KryszewskiS001993,BuhmannSY042005})
\begin{equation}
\label{L38}
V^\mathrm{(at)}(z)
=\frac{\hbar\mu_{0}}{8\pi^2
}
\int_{0}^{\infty}\D\xi\,
\xi^2
\alpha(i\xi)\int_{0}^{\infty}
\D q\,
\frac{q}{\kappa}\,
e^{-2 \kappa z}
\left[
r_{1-}^{s}
-r_{1-}^{p}
 \left(1+\frac{2q^2}
 {\xi^2/c^2}
 \right)
\right]
\end{equation}
($\kappa^2$ $\!=$ $\!\xi^2/c^2$ $\!+$ $\!q^2$), where
the atom is situated at some position $z$ $\!>$ $\!0$, and the
magnetodielectric wall %
%proof:
%[
(which may have
a (1D) internal structure,
%proof:
%e.g.,
for example, %
a planarly layered one%
%proof:
%]
)
extends from some negative $z$-value up to $z$ $\!=$ $\!0$.
Note that the effect of the wall is fully
described in terms of the
(generalized) reflection coefficients
$r_{1-}^{s}$ and $r_{1-}^{p}$,
both of which are functions of
the imaginary frequency $i\xi$
and the transverse wave vector projection $q$
($s,p$, polarization indices).

Two comments on equations
(\ref{L6}) and (\ref{L38}) seem to be advisable.
First, the neglect of level broadening
might suggest that $\alpha(\omega)$
has poles on the real frequency axis, such as
\begin{equation}
\label{L5a-0}
\alpha(\omega)
\sim
\sum_{k}
\frac{\Omega_{k}^2}{\omega_{k}^2-\omega^2}\,.
\end{equation}
In fact, equation (\ref{L5a-0}) has to be understood as
\begin{equation}
\label{L5a}
\alpha(\omega)
\sim
\lim_{\gamma\to 0+}
\sum_{k}
\frac{\Omega_{k}^2}
{\omega_{k}^2-\omega^2-i\gamma\omega}\,,
\end{equation}
where the limit prescription
$\gamma$ $\!\to$ $\! 0+$ reminds one of
the proper response function properties \cite{LanLif} of $\alpha(\omega)$.
Throughout this paper it will not be necessary to make use of any
particular form of the polarizability. Second, although
the (magnetodielectric) bodies can be quite arbitrary, it
is important that the atom under study is an isolated one,
i.e., equations (\ref{L6}) and (\ref{L38}) do not apply
to atoms in matter.
Needless to say that a convincing consideration of atoms
in matter
must include local field corrections.
Correspondingly, the reflection
coefficients in equation (\ref{L38}) refer to the
reflection of waves being
incident on the wall from free space.

%%%%%%%%%%%%%%%%%%%%%%%%%%%%%%%%%%%%%%%%%%%%%%%%%%%%%%%%%%%%%%%

\section{Casimir force}
\label{sec3}

Let us now
%proof:
%ask for
consider
the Casimir force acting on a macroscopic body
in the presence of other bodies.
In the zero-temperature limit, it is just the
ground-state expectation value of the
Lorentz force acting on the charges and currents which constitute
the body on the level of macroscopic electrodynamics, i.e.,
the charge density $\hat{\rho}(\mathbf{r})$ can be given by
\begin{equation}
\label{L40}
\hat{\rho}(\mathbf{r})=\int_{0}^\infty
\D\omega\,
\fo{\rho}(\mathbf{r},\omega) + \mathrm{H.\,c.}
\end{equation}
and the current density
$\hat{\mathbf{j}}(\mathbf{r})$ accordingly, with
\begin{equation}
\label{L40a}
\fo{\rho}(\mathbf{r},\omega) =
-\varepsilon_{0}
\Nabla\cdot\{[\varepsilon(\mathbf{r},\omega)-1]\fo{E}(\mathbf{r},\omega)\}
+ (i\omega)^{-1} \Nabla\cdot\fo{j}_\mathrm{N}(\mathbf{r},\omega)
\end{equation}
and
\begin{equation}
\label{L40b}
\fl
\fo{j}(\mathbf{r},\omega)
=
-i\omega \varepsilon_{0}
[\varepsilon(\mathbf{r},\omega)-1]
\fo{E}(\mathbf{r},\omega)
+ \Nabla\times \{
\mu_{0}^{-1}
[1-\mu^{-1}(\mathbf{r},\omega)]
\fo{B}(\mathbf{r},\omega)\}
+\fo{j}_\mathrm{N}(\mathbf{r},\omega)
\end{equation}
(for details, see
%proof:
%reference
%
\cite{RaabeC012005}).
Here, $\fo{j}_{\mathrm{N}}(\mathbf{r},\omega)$
is the current density that acts as a Langevin noise source
in the operator Maxwell equations,
$\varepsilon(\mathbf{r},\omega)$ and
$\mu(\mathbf{r},\omega)$
are the permittivity and permeability,
respectively, and
$\fo{E}(\mathbf{r},\omega)$ and $\fo{B}(\mathbf{r},\omega)$
are the (positive) frequency parts of the
electric field and the induction field, respectively,
\begin{equation}
\label{L43}
\fo{E}(\mathbf{r},\omega)
=i\mu_{0}\omega\int \D^3r'\,
\tensor{G}(\mathbf{r,r'},\omega)
\cdot
\fo{j}_{\mathrm{N}}(\mathbf{r'},\omega),
\end{equation}
\begin{equation}
\label{L44}
\fo{B}(\mathbf{r},\omega)
=\mu_{0}\Nabla\times \int \D^3r'\,
\tensor{G}(\mathbf{r,r'},\omega)
\cdot
\fo{j}_{\mathrm{N}}(\mathbf{r'},\omega).
\end{equation}
According to
%proof:
%reference
%
\cite{RaabeC012005},
the Casimir force on a magnetodielectric body of volume $V$
can then be expressed in terms of the Casimir stress as
a surface integral,
\begin{equation}
\label{L44-1}
\mathbf{F}=\int_{\partial V}
\D\mathbf{a}
\cdot \tensor{T}(\mathbf{r,r}),
\end{equation}
where
\begin{equation}
\label{L12}
\tensor{T}(\mathbf{r,r})=
\lim_{\mathbf{r}'\to\mathbf{r}}\left[
\tensor{\theta}(\mathbf{r,r'})
- {\textstyle\frac{1}{2}} \tensor{1}
{\rm Tr}\,\tensor{\theta}(\mathbf{r,r'})
\right],
\end{equation}
\begin{equation}
\label{L13}
\tensor{\theta}(\mathbf{r,r'})
=-\frac{\hbar}{\pi}\int_{0}^{\infty} \D\xi\,
\left[\frac{\xi^2}{c^2}\,\tensor{G}^\mathrm{(S)}(\mathbf{r,r'},i\xi)
+\Nabla\times
\tensor{G}^\mathrm{(S)}(\mathbf{r,r'},i\xi)\times\Lnabla{'}
\right]\!,
\end{equation}
with the body under study being taken into account in the
definition of the scattering Green tensor. It is worth noting that
equations (\ref{L12}) and (\ref{L13}) also
apply if the interspace between the bodies is not empty but
also filled with magnetodielectric matter
(which has to be homogeneous at least in some small
neighbourhood of the body under consideration). In this case
the Casimir force is expected to be diminished
as compared to the case where the surrounding matter
is absent, because of the screening effect of the matter.

Let us apply equation (\ref{L12})
%proof:
%[
(%
together with equation (\ref{L13})%
%proof:
%]
)
to the $j$th (homogeneous) layer of a planar magnetodielectric
multi-layer structure. For such systems the Green tensor is well
known \cite{TomasM031995,ChewBook}, leading to
($0$ $\!<$ $\!z$ $\!<$ $\!d_j$; $d_j$, thickness of the layer)
\begin{equation}
\label{L201}
T_{zz}(\mathbf{r,r})=
\frac{\hbar}{8\pi^2}\int_{0}^{\infty}\!\!\D\xi\,
\int_{0}^{\infty} \D q\,q\,\frac{\mu_{j}(i\xi)}
{i\beta_{j}(i\xi,q)}\, g_j(z,i\xi,q),
\end{equation}
where
\begin{eqnarray}
\label{L202}
g_j(z,\omega,q)
&=& 2\bigl[\beta_{j}^2 (1+n^{-2}_{j})-q^2
(1-n^{-2}_{j})\bigr]
D_{js}^{-1}r_{j+}^{s}r_{j-}^{s}e^{2i\beta_{j}d_{j}}
\nonumber\\
&& +2\bigl[\beta_{j}^2 (1+n^{-2}_{j})
+q^2 (1-n^{-2}_{j})\bigr]
D_{jp}^{-1}r_{j+}^{p}r_{j-}^{p}e^{2i\beta_{j}d_{j}}
\nonumber\\
&& - (\beta_{j}^2+q^2)(1-n^{-2}_{j})
D_{js}^{-1}\bigl[r_{j-}^{s}
e^{2i\beta_{j} z}+r_{j+}^{s}
e^{2i\beta_{j}(d_{j}-z)}\bigr]
\nonumber\\
&& + (\beta_{j}^2+q^2)(1-n^{-2}_{j})
D_{jp}^{-1}
\bigl[r_{j-}^{p}e^{2i\beta_{j} z}+r_{j+}^{p}
e^{2i\beta_{j}(d_{j}-z)}\bigr],
\end{eqnarray}
with
\begin{eqnarray}
\label{L203}
n^2_{j}
=n^2_{j}(\omega)
=\varepsilon_{j}(\omega)\mu_{j}(\omega),\\
\label{L204}
\beta_{j}=\beta_{j}(\omega,q)=(\omega^2n_{j}^2/c^2
-q^2)^{1/2},\\
\label{L205}
D_{j\sigma}=D_{j\sigma}(\omega,q)=
1-r_{j+}^{\sigma} r_{j-}^{\sigma} e^{2i\beta_{j}d_{j}},
\end{eqnarray}
and $r_{j\pm}^{\sigma}$ $\!=$ $\!r_{j\pm}^{\sigma}(\omega,q)$ being
the generalized reflection coefficients associated with
the $j$th layer ($\sigma$ $\!=$ $\!s,p$).
Since $-i\beta_{j}$ is purely
real and nonnegative at imaginary frequencies,
we will use the notation \mbox{$\kappa_{j}$ $\!=$
$\!-i\beta_{j}(i\xi,q)$} in the remainder of the paper.

%%%%%%%%%%%%%%%%%%%%%%%%%%%%%%%%%%%%%%%%%%%%%%%%%%%%%%%%%%%%%%
%%%%%%%%%%%%%%%%%%%%%%%%%%%%%%%%%%%%%%%%%%%%%%%%%%%%%%%%%%%%%%

\section{Unscreened versus screened Casimir-Polder force}
\label{sec4}
\begin{figure}[htb]
\centering
\includegraphics[width=8cm,angle=0]{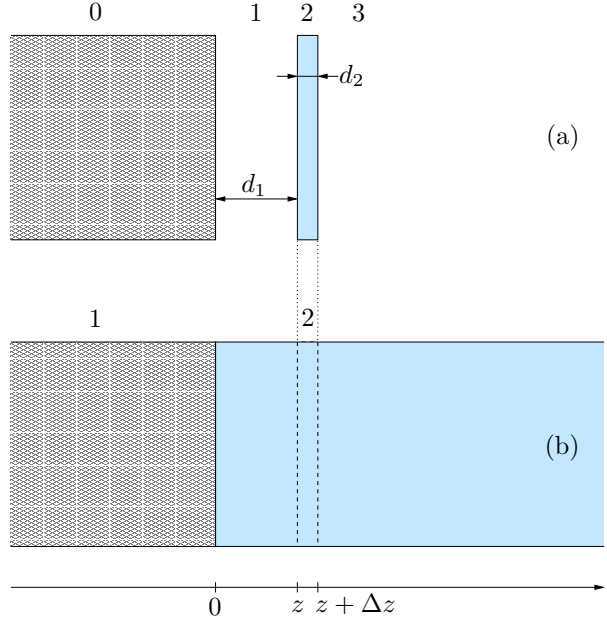}
\caption{\label{Fig1}
A homogeneous plate in front of a planar magnetodielectric
wall (a); the plate is part of matter that fills
the space in front of the wall (b).
%proof:
\\[1ex]
(This figure is in colour only in the electronic version)
}
\end{figure}

Equation (\ref{L44-1}) together with equations (\ref{L12})
and (\ref{L13}) contains the unscreened
CP force that acts on isolated atoms as limiting case.
Moreover, it enables one to calculate also the
screened CP force acting on atoms that are
constituents of matter. To illustrate this, let us consider planar
systems as sketched in figure \ref{Fig1} and begin with
the unscreened CP force.

\subsection{
%proof:
%Casimir-Polder
Casimir--Polder
force on isolated atoms}
\label{sec4.2}
For this purpose we first calculate the Casimir force
acting on a (homogeneous) plate of thickness
\mbox{$d_2$ $\!=$ $\!\Delta z$} in
front of a planar wall according to the four-layer system in
figure \ref{Fig1}(a), with the regions
1 and 3 being empty. In this case we have to set
$n_{1}$ $\!=$ $\!n_{3}$ $\!=$ $\!1$ and $d_{3}$ $\!\to$ $\!\infty$
(and hence \mbox{$r_{3+}^{\sigma}$ $\!\to$ $\!0$)} in
equations (\ref{L201}) and (\ref{L202}), which in particular implies
that the stress in the empty-space region 1 is independent of position
and vanishes in the semi-infinite empty-space region 3.
According to equations (\ref{L44-1}) and (\ref{L201}), the
total Casimir force (per transverse unit area) acting on
the plate (layer $2$) can then be written as
[$\kappa_{1}$ $\!=$ $\!\kappa_{3}$ $\!=$
$\!\kappa$ $\!=$ $\!(\xi^2/c^2+q^2)^{1/2}$]
\begin{equation}
\label{L24}
F=\frac{\hbar}{8\pi^2}\int_{0}^{\infty}\D\xi\int_{0}^{\infty}
\D q\,\frac{q}{\kappa}\, g_{1}(d_{1},i\xi,q),
\end{equation}
where
\begin{equation}
\label{L25}
\fl
g_{1}(d_{1},i\xi,q)
=
-4\kappa^2
\sum_{\sigma=s,p}\frac{r_{1+}^{\sigma}r_{1-}^{\sigma}
e^{-2\kappa d_{1}}}{1-r_{1+}^{\sigma}r_{1-}^{\sigma}
e^{-2\kappa d_{1}}}
=-4\kappa^2
\sum_{\sigma=s,p}\sum_{m=1}^{\infty}[r_{1+}^{\sigma}r_{1-}^{\sigma}
e^{-2\kappa d_{1}}]^{m}.
\end{equation}
Since the reflection coefficients $r_{1-}^{\sigma}$ (containing
the details of the wall structure) are independent of
the properties of the plate, they
need not be further specified.
In
%proof:
the
case of a homogeneous plate,
the reflection coefficients $r_{1+}^{\sigma}$
read
\begin{equation}
\label{L26}
r_{1+}^{\sigma}
=\frac{r_{1/2}^{\sigma}\,(1-e^{-2\kappa_{2}d_{2}})}
{1-r_{1/2}^{\sigma\,2}e^{-2\kappa_{2}d_{2}}}
\qquad (\sigma = s,p),
\end{equation}
where $r_{1/2}^{\sigma}$ are the usual single-interface
(Fresnel) amplitudes,
\begin{equation}
\label{L27}
r_{1/2}^{s}=
\frac{\kappa\mu_{2}-\kappa_{2}}
{\kappa\mu_{2}+\kappa_{2}}\,,
\qquad
r_{1/2}^{p}
=\frac{\kappa\varepsilon_{2}-\kappa_{2}}
{\kappa\varepsilon_{2}+\kappa_{2}}\,.
\end{equation}

To recover the ordinary (unscreened) CP force,
let us consider a  nonmagnetic plate ($\mu_{2}$ $\!=$ $\!1$)
consisting of weakly dielectric material
($|\varepsilon_{2}$ $\!-$ $\!1|$ $\!\ll$ $\!1$)
and expand to first order in $\varepsilon_{2}$ $\!-$ $\!1$.
In this approximation, we may set
\begin{equation}
\label{L28a}
\frac{\kappa_{2}}{\kappa} =1+\frac{2\xi^2}{\kappa^2
c^2}\,(\varepsilon_{2}-1)
\end{equation}
and equations (\ref{L27}) approximate to
\begin{equation}
\label{L29}
r_{1/2}^{s}
=-
(\varepsilon_{2}-1)
\,\frac{\xi^2}
{4\kappa^2 c^2}\,,
\end{equation}
\begin{equation}
r_{1/2}^{p}=\frac{
\varepsilon_{2}-1
}{2}\,
\left(1-
\frac{\xi^2}
{2\kappa^2 c^2}
\right),
\end{equation}
so that equations (\ref{L26}) read
\begin{equation}
\label{L31a}
r_{1+}^{s}
=-
(\varepsilon_{2}-1)
\,\frac{\xi^2}
{4\kappa^2 c^2}
\left(1-e^{-2\kappa
%d_{2}
\Delta z
}\right),
\end{equation}
\begin{equation}
\label{L31b}
r_{1+}^{p}=
\frac{
\varepsilon_{2}-1
}
{2}
\left(1-
\frac{\xi^2}
{2\kappa^2 c^2}
\right)\left(1-e^{-2\kappa
\Delta z
}\right)
.
\end{equation}
Inserting equations (\ref{L31a}) and  (\ref{L31b}) in
equation (\ref{L25}), we see that $g_{1}(d_{1},i\xi,q)$
approximates to
\begin{eqnarray}
\label{L32}
\fl
g_{1}(d_{1},i\xi,q)=-4\kappa^2
\sum_{\sigma=s,p}
r_{1+}^{\sigma}r_{1-}^{\sigma}
e^{-2\kappa d_{1}}
\nonumber\\\lo
=
(\varepsilon_{2}-1)
\,\kappa^2 e^{-2\kappa d_{1}}
\left(1-e^{-2\kappa
\Delta z
}\right)
\left[
r_{1-}^{s}
 \left(1-\frac{q^2}{\kappa^2}
 \right)
-r_{1-}^{p}
 \left(1+\frac{q^2}{\kappa^2}
 \right)
\right]
.
\end{eqnarray}
Note that it solely results
from the $m$ $\!=$ $\!1$ (%
%proof:
%``
`single round-trip'%
%''
) term
%proof:
%[
(%
in the second, expanded form of equation (\ref{L25})%
%]
). Substitution of equation (\ref{L32})
into equation (\ref{L24}) yields the
Casimir force (per transverse unit area)
to first order in $\varepsilon_{2}$ $\!-$ $\!1$:
\begin{eqnarray}
\label{L24a}
\fl
F=\frac{\hbar}{8\pi^2}\int_{0}^{\infty}\D\xi\,
(\varepsilon_{2}-1)
\nonumber\\ \times
\int_{0}^{\infty}
\D q\,q\kappa
 e^{-2\kappa d_{1}}
\left(1-e^{-2\kappa
\Delta z
}\right)
\left[
r_{1-}^{s}
 \left(1-\frac{q^2}{\kappa^2}
 \right)
-r_{1-}^{p}
 \left(1+\frac{q^2}{\kappa^2}
 \right)
\right].
\end{eqnarray}
It is not difficult to prove that equation (\ref{L24a})
can be rewritten as
\begin{equation}
\label{L24a-1}
F = \int_{d_1}^{d_1+\Delta z} \D z\, f(z),
\end{equation}
where the force density $f(z)$ can be derived from
a potential $V(z)$ as follows:
\begin{equation}
\label{L35a}
f(z) = -
\frac{\partial V(z)}{\partial z}
\,,
\end{equation}
\begin{equation}
\label{L37a}
\fl
V(z)=\frac{\hbar
}
{8\pi^2
}
\int_{0}^{\infty}
\D\xi\,
(\varepsilon_{2}-1)
\int_{0}^{\infty}
\D q\,q\kappa
 e^{-2\kappa z}
\left[
r_{1-}^{s}
 \left(1-\frac{q^2}{\kappa^2}
 \right)
-r_{1-}^{p}
 \left(1+\frac{q^2}{\kappa^2}
 \right)
\right].
\end{equation}
Let us suppose that the plate consists of atom-like basic
constituents of polarizability $\alpha(i\xi)$
and $\eta$ is the (constant) number density
of the atoms. Using the relation
\begin{equation}
\label{L25c}
\varepsilon_{2}(i\xi)-1
=\frac{\eta}
{\varepsilon_{0}}
\, \alpha(i\xi),
\end{equation}
which is valid in the case of
weakly dielectric material,
from inspection of equations (\ref{L24a-1})--(\ref{L37a})
we see that
\begin{equation}
\label{L25c-1}
F^{({\rm at})}(z) = \eta^{-1}f(z)
\end{equation}
can be regarded as the force acting on an atom at position $z$
in the plate, where the associated potential reads
\begin{equation}
\label{L37}
\fl
V^\mathrm{(at)}
(z)
=\frac{\hbar}{8\pi^2
\varepsilon_{0}
}
\int_{0}^{\infty}\D\xi\,
\alpha(i\xi)
\int_{0}^{\infty}
\D q\, q \kappa e^{-2 \kappa z}
\left[
r_{1-}^{s}
 \left(1-\frac{q^2}{\kappa^2}
 \right)
-r_{1-}^{p}
 \left(1+\frac{q^2}{\kappa^2}
 \right)
\right].
\end{equation}
It is straightforwardly checked that equation (\ref{L37}) is
identical with equation (\ref{L38}), i.e., with the standard CP
potential. Hence, $F^{({\rm at})}(z)$ is nothing but the
unscreened CP force that acts on a single (ground-state)
atom at position $z$ in front of the wall.
Clearly, this interpretation presupposes that $\alpha(i\xi)$
is really the polarizability of a single atom. Otherwise
$\alpha(i\xi)$ and $\eta$ are rather formal quantities
defined by equation (\ref{L25c}), so that
the introduction of
$F^{({\rm at})}(z)$ is also rather formal.
Although equations
(\ref{L25c-1}) and (\ref{L37}) are of course correct, equations
(\ref{L35a}) and (\ref{L37a}) may be more appropriate in this case.
%
%%%%%%%%%%%%%%%%%%%%%%%%%%%%%%%%%%%%%%%%%%%%%%%%%%%%%%%%%%%%%%%%%%%%%%%

\subsection{Screened Casimir-Polder force on medium atoms}
\label{sec4.1}
Let us now consider the case where the plate is part of
matter that fills the space in front of the wall according
to the two-layer system in figure \ref{Fig1}(b). In this case
from equation (\ref{L202}) it follows that ($d_2$ $\!\to$ $\!\infty$
and hence $r^\sigma_{2+}$ $\!=$ $\!0$)
\begin{equation}
\label{L100}
\fl
g_{2}(z+\Delta
z,\omega,q)-g_{2}(z,\omega,q)=
e^{-2\kappa_{2} z}
(q^2-\kappa_{2}^2)
(1-n_{2}^{-2})
(r_{2-}^{p}-r_{2-}^{s})
\left(e^{-2\kappa_{2}\Delta z}-1\right),
\end{equation}
so that, according to equations (\ref{L44-1}) and (\ref{L201})
the force (per transverse unit area) on the plate reads
\begin{equation}
\label{L101}
F=\frac{\hbar}{8\pi^2}\!\int_{0}^{\infty}\D\xi\,
\frac{\xi^2}{c^2}\,\mu_{2}
(n_{2}^2-1)
\!\int_{0}^{\infty}\D q\, \frac{q}{\kappa_{2}}
\,
e^{-2\kappa_{2} z}
(r_{2-}^{p}-
r_{2-}^{s})
\left(e^{-2\kappa_{2}\Delta z}-1\right).
\end{equation}

Now we again focus on nonmagnetic
and weakly dielectric matter, i.e.,
$\mu_{2}$ $\!=$ $\!1$,
\mbox{$|\varepsilon_{2}$ $\!-$ $\!1|$ $\!\ll$ $\!1$}.
It is not difficult to see that to
first order in $\varepsilon_{2}$ $\!-$ $\!1$
equation (\ref{L101}) yields
\begin{equation}
\label{L102}
F=\frac{\hbar}{8\pi^2}\!\int_{0}^{\infty}\D\xi\,
\frac{\xi^2}{c^2}\,
(\varepsilon_{2}-1)
\!\int_{0}^{\infty}\D q\,\frac{q}{\kappa}
\,
e^{-2\kappa z}
(r_{2-}^{p}-
r_{2-}^{s})
\left(e^{-2\kappa\Delta z}-1\right),
\end{equation}
where the quantities $r_{2-}^{\sigma}$
must be computed for
$\varepsilon_{2}$ $\!=$ $\!1$ and $\kappa_{2}$ $\!=$ $\!\kappa$.
In other words, they are the reflection coefficients
corresponding to the case where
the half-space on the right-hand side
of the wall is empty, and thus they
agree with the reflection coefficients
$r_{1-}^{\sigma}$ in equation (\ref{L24a}).
Obviously, equation (\ref{L102}) can be written
in the form of equation (\ref{L24a-1})
(with $d_{1}\mapsto z$),
where the force density $f(z)$ can be derived,
according to equation (\ref{L35a}), from a potential $V(z)$
which now reads
\begin{equation}
\label{L103a}
V(z)=-\frac{\hbar}{8\pi^2}\,
\int_{0}^{\infty}\D\xi\,
\frac{\xi^2}{c^2}\,
(\varepsilon_{2}-1)
\int_{0}^{\infty}\D q\,
\frac{q}{\kappa}
\,e^{-2\kappa z}
(r_{2-}^{p}-
r_{2-}^{s})
\end{equation}
instead of equation (\ref{L37a}).

Applying equation (\ref{L25c}), we can again use equation
(\ref{L25c-1}) to introduce the force $F^{({\rm at})}(z)$ acting
on an atom at position $z$ in the plate, where the
potential from which the force can be derived is given by
\begin{equation}
\label{L105b}
V^\mathrm{(at)}(z)
=-\frac{\hbar\mu_{0}}{8\pi^2}\,
\int_{0}^{\infty}\D\xi\,
\xi^2\,\alpha(i\xi)
\int_{0}^{\infty}\frac{\D q\,q}{\kappa}\,
(r_{2-}^{p}-r_{2-}^{s})
e^{-2\kappa z}
\end{equation}
instead of equation (\ref{L37}).
Hence, equation (\ref{L105b}) can be interpreted as the potential
that a matter atom is subject to when
the presence of the surrounding
atoms of the (weakly dielectric) matter
is taken into account,
thus being the screened single-atom CP potential.
It differs from
the unscreened CP potential (\ref{L38})
%proof:
%[
(%
or, equivalently, (\ref{L37})%
%proof:
%]
)
in the  $p$-polarization contributions.
The result is in agreement with the one recently found
in
%proof:
%reference
%
\cite{TomasM022005}, in which
equation (\ref{L44-1}) together with equation (\ref{L201})
is applied to a multi-plate cavity-like system
and it is shown that the Casimir force acting on a plate
embedded in such a system can be decomposed into two
parts, where one part can be regarded as screened force.

%%%%%%%%%%%%%%%%%%%%%%%%%%%%%%%%%%%%%%%%%%%%%%%%%%%%%%%%%%%%%%%%%

\section{Discussion and Summary}
\label{sec5}
It is not surprising that the unscreened CP potential
%proof:
%[
(%
equation (\ref{L38}) or, equivalently, equation
(\ref{L37})%
%]
)
must differ from the screened one
%proof:
%[
(%
equation (\ref{L105b})%
%],
),
as can be seen from the respective
physical meaning of the force (per transverse unit area)
$\D F$ $\!=$ $\!f(z)\D z$ that acts on a slice of
infinitesimal thickness $\D z$
%proof:
%[cf.~equation (\ref{L24a-1})].
(cf~equation (\ref{L24a-1})).
To obtain the unscreened CP force, a slice in the otherwise
empty right half-space is considered
%proof:
%[
(%
see figure \ref{Fig1}(a)%
%]
),
leading -- if equation (\ref{L25c}) holds -- to the force
on a single atom in front of the wall.
%proof: new paragraph

In contrast, the screened CP force is obtained if the slice is
unavoidably a part of the medium that fills the right half-space in
figure \ref{Fig1}(b), leading to the force on a medium atom. From
a microscopical point of view, this force does not only result
from the van der Waals forces between the medium
atom under consideration
and the atoms of the wall, but also between the medium
atom and the other medium atoms.
Since there are
%proof:
%much more
many more
other medium atoms to the right rather than to
the left of the medium atom under consideration, there will be a net
effect, which is expected to diminish the force as
compared to the single-atom case.

To illustrate this
screening effect,
let us compare the unscreened potential
%proof:
%[
(%
equation (\ref{L37})%
%]
)
and the screened potential
%proof:
%[
(%
equation (\ref{L105b})%
%]
)
in the idealized limit where the wall can be
regarded as a perfectly reflecting mirror such that
$r_{1-}^{p}$ $\!=$ $\! 1$,
\mbox{$r_{1-}^{s}$ $\!=$ $\!-1$}.
In this case equation (\ref{L37}) greatly simplifies
(the expression in the square bracket equals $-2$),
leading to the unscreened potential in the form of
\begin{equation}
\label{L33}
V^\mathrm{(at)}(z)
=-\frac{\hbar c}{64\pi^2\varepsilon_{0}
}
\frac{1}{z^4}\int_{0}^{\infty} \D y\,
\alpha\!\left(\frac{icy}{2z}\right)h
(y),
\end{equation}
where the dimensionless function $h(y)$ is defined by
\begin{equation}
\label{L24c}
h(y)
=\int_{y}^{\infty}\D x\, x^2 e^{-x}=(y^2+2y+2)e^{-y}.
\end{equation}
In particular, in the large-distance
limit equation (\ref{L33}) reduces to
Casimir's and Polder's well-known formula
\cite{CasimirHBG021948}
$\bigl[\int_{0}^{\infty} \D y \,h(y)$ $\!=$ $\!6\bigr]$
\begin{equation}
\label{L34}
V^\mathrm{(at)}(z)
=
- \frac{3\hbar c\alpha(0)}{32\pi^2\varepsilon_{0}}\,
\frac{1}{z^4}
\,.
\end{equation}
Correspondingly, setting
$r_{2-}^{p}$ $\!=$ $\!1$ and $r_{2-}^{s}$
$\!=$ $\! -1$ in equation (\ref{L105b}),
we may write the screened potential in the form of
equation (\ref{L33}), where the function $h(y)$ changes to
\begin{equation}
\label{L24d-1}
h(y)
=y^2 \int_{y}^{\infty}\D x\, e^{-x}=y^2 e^{-y}.
\end{equation}
Since now $\int_{0}^{\infty} \D y \,h(y)$ $\!=$ $\!2$, it
follows that in the large distance-limit
the screened potential is one third of the unscreened one,
\begin{equation}
\label{L34-1}
V^\mathrm{(at)}(z)
=
- \frac{\hbar c\alpha(0)}{32\pi^2\varepsilon_{0}}\,
\frac{1}{z^4}\,,
\end{equation}
provided that $\alpha(0)$ is the same in both cases.
The result clearly shows that the screening effect
can be fairly large under certain conditions.
In the short-distance limit, the
asymptotic behaviour of the unscreened CP potential is commonly
obtained by approximately setting \mbox{$c$ $\!\to$ $\!\infty$},
implying $\kappa$ $\!\simeq$ $\!q$
(non-retarded approximation), thus the leading term behaves like
$z^{-3}$. It is not difficult to see that this term is missing
in the screened potential, and that the same approximation would lead
to a $z^{-1}$ distance law. It is, however, more than questionable
whether this approximation is reasonable in the case of a medium atom.
Moreover, when the atom can be no longer regarded as being
surrounded by sufficiently many other atoms of the same
kind, then the macroscopic description leading to
equation (\ref{L105b}) fails. Hence, even if the strict short-distance limit of
Eq.~(\ref{L105b}) could be found by more elaborate methods, its meaning
might be severly limited.

{F}rom the above we can conclude that electromagnetic vacuum forces
acting on micro-objects that consist of collections of atomic
constituents should be preferably calculated as
Casimir forces, by assigning appropriately chosen permittivities
and/or permeabilities to the micro-objects.
This approach ensures that screening effects are properly taken into account.
In particular, in the case of a weakly magnetodielectric
object the total force can be obtained,
%proof:
%in
to
leading order, by
superimposing screened CP forces acting on the atomic constituents.
With increasing strength of the magnetodielectric properties,
higher-order corrections (not considered in this paper)
must be included in the calculation.
This can be done in a systematic fashion, by
starting from the exact formula for the Casimir force and
expanding to higher powers in the electric and/or
magnetic susceptibility.

In summary, we have
studied relations between the Casimir force acting on
a macroscopic body and the CP force acting on an atom, with
special emphasis on planar structures.
We have shown that the exact formula for the
Casimir force contains as
%proof:
a
special case the
ordinary CP force acting
on a single atom in front of a wall.
%proof:
%Further
Further, %
we have shown that
the exact formula can also be used to calculate
the CP force acting on an atom
that is
%proof:
a
constituent
of bulk material that fills the half-space in front of the wall.
In this case the surrounding atoms give rise
to a screening effect that diminishes the
CP force compared with the force that acts
on an isolated atom.

\ack
We thank
Stefan Scheel for discussions.
%proof:
%C.~R.
CR %
is grateful for being granted a Th\"u\-rin\-ger
Lan\-des\-gradu\-ier\-ten\-sti\-pen\-dium.

%%%%%%%%%%%%%%%%%%%%%%%%%%%%%%%%%%%%%%%%%%%%%%%%%%%%%%%%%%%%%%%%%%%%%%
\section*{References}
%

%%%%%%%%%%%%%%%%%%%%%%%%%%%%%%%%%%%%%%%%%%%%%%%
\end{document}